\documentclass[a4paper,11pt]{article}
\pdfoutput=1 

\usepackage{jheppub} 

\usepackage[T1]{fontenc} 
\usepackage{slashed}

\preprint{arXiv:1303.1782 [hep-th]}

\title{Non-Associative Geometry and the Spectral Action Principle}
  
\author{Shane Farnsworth}
\author{and Latham Boyle}
\affiliation{Perimeter Institute for Theoretical Physics, \\ 
Waterloo, Ontario N2L 2Y5, Canada}  

\abstract{Chamseddine and Connes have argued that the action for
  Einstein gravity, coupled to the $SU(3)\times SU(2)\times U(1)$
  standard model of particle physics, may be elegantly recast as the
  ``spectral action'' on a certain ``non-commutative geometry.''  In
  this paper, we show how this formalism may be extended to
  ``non-associative geometries,'' and explain the motivations for
  doing so.  As a guiding illustration, we present the simplest
  non-associative geometry (based on the octonions) and evaluate its
  spectral action: it describes Einstein gravity coupled to a $G_2$
  gauge theory, with 8 Dirac fermions (which transform as a singlet
  and a septuplet under $G_2$).  This is just the simplest example: in
  a forthcoming paper we show how to construct more realistic models
  that include Higgs fields, spontaneous symmetry breaking and fermion
  masses.}

\begin{document}
\maketitle


\section{Introduction and Motivation}
\label{Introduction}
Despite the success of the standard model of particle physics (SM),
the set of gauge theories consistent with current experimental
constraints remains very large. It is a striking fact therefore that
the SM is also a member of a much more restricted set of gauge
theories -- those that may be reinterpreted as arising from non-commutative
geometry (NCG) in the sense of Connes\cite{Connes199129,
  Connes:1995tu, Connes:1996gi, Chamseddine:1996zu,
  Chamseddine:1996rw, Chamseddine:1991qh, Barrett:2006qq,
  Connes:2006qv, Chamseddine:2006ep, Chamseddine200838,
  Chamseddine:2007ia, Chamseddine:2010ud, Chamseddine:2012sw}. NCG was
developed over the past few decades as a generalization of Riemannian
geometry. For an in depth review of NCG and the standard model
embedding, aimed at physicists, see {\it e.g.}\ \cite{Schucker:2001aa,
  vandenDungen:2012ky}.
    
The axioms of NCG place severe restrictions on the allowed symmetries
and particle content which may be geometrically modeled in this way
(see {\it e.g.}\ \cite{Chamseddine200838, Chamseddine:2007ia, Krajewski:1996se}).
One might hope therefore to use the framework of NCG to explore beyond
the standard model of physics.  In doing so, however, one should think
carefully about which constraints are truly natural or intrinsic to
this approach, and which are imposed artificially.  In this
paper we argue that there are good reasons to consider relaxing the
associativity restrictions in the NCG formalism, and we take some
of the first steps towards such a generalization.

There are two main motivations for a non-associative
generalization of the NCG framework: a more general mathematical one
and a more specific physical one.

Let us start with the more general mathematical motivation.  The
fundamental point is that, in the ordinary approach to physics, the
basic input is a {\it symmetry group}.  By contrast, in the spectral
approach, the fundamental input is an {\it algebra}, and the symmetry
group then emerges as the automorphism group of that algebra.
Symmetry groups are associative by nature, but algebras are not.  Just
as some of the most beautiful and important groups are noncommutative,
some of the most beautiful and important algebras (including Lie
algebras, Jordan algebras and the Octonions) are nonassociative.  Just
as it would be unnatural to restrict our attention to commutative
groups (as physicists originally did in studying gauge theory, prior
to Yang-Mills), it is unnatural to restrict our attention to associative
algebras.  In either case, imposing such an unnatural restriction
likely amounts to blinding ourselves to something essential that the
formalism is trying to tell us.  From this standpoint, our task is to
formulate the spectral approach to physics in such a way that the
extension to nonassociative algebras becomes obvious and natural.

Next we turn to the more specific physical motivation.  Although we
would like to use the framework of non-commutative geometry to explore
beyond the standard model of particle physics, many of the most
interesting extensions are out of reach of the associative formalism.
As a specific example, in order to reformulate the most successful
Grand Unified Theories (GUTs) -- {\it e.g.}\ those based on $SU(5)$,
$SO(10)$ and $E_{6}$ -- in terms of the spectral action, we are forced
to use nonassociative input algebras.  To appreciate this point, first
note that the representation theory of associative $\ast$-algebras is
much more restricted than the representation theory of Lie groups
\cite{Krajewski:1996se}: Lie groups (like $SU(5)$) have an infinite
number of irreps, but associative algebras (like the corresponding
$\ast$-algebra $M_{5}(\mathbb{C})$ of $5\times5$ complex matrices,
whose automorphism group is $SU(5)$) only have a small finite number.  In
particular, if we ask whether key fermionic representations needed in
GUT model building -- such as the ${\bf 10}$ of $SU(5)$, the ${\bf
  16}$ of $SO(10)$, or the ${\bf 27}$ of $E_{6}$ -- are available as
the irreps of algebras with the correct corresponding automorphism
groups, the answer is ``no'' for associative algebras, and ``yes'' for
nonassociative algebras.  Furthermore, if we ask whether the
exceptional groups (including $E_{6}$, which is of particular interest
for GUT model building, and $E_{8}$, which is of particular interest
in connection with string theory) appear as the automorphism groups of
corresponding algebras, again the answer is ``no'' for associative
algebras and ``yes'' for nonassociative algebras.

With these motivations in mind, in Section \ref{NAGformalism} we
consider the aspects of NCG which must be generalized or recast in
order to accomodate non-associativity, and we take some of the first
steps towards formulating non-associative geometry.  Keeping in mind
the application to physics, we pay particular attention to a class of
geometries that we call `almost-associative'.  These are constructed
by taking the product of an ordinary (commutative, associative,
infinite-dimensional) smooth Riemannian spectral triple on the one
hand, and a finite-dimensional non-associative spectral triple on the
other.

One of our key physics results (explained in Sections
\ref{NAGformalism} and \ref{sec4}) is the following: from the spectral
action on an `almost-associative' geometry, one obtains the action for
an {\it ordinary} gauge theory, coupled to {\it ordinary} Einstein
gravity, built from {\it ordinary} scalar, spinor, gauge and
metric fields, and living on {\it ordinary} spacetime.  In other
words, the reader might worry that perhaps if the underlying algebra
is non-associative, then the spectral action will produce some sort of
exotic non-associative theory -- {\it e.g.}\ built from some sort of
exotic non-associative gauge fields, or living on some sort of exotic
non-associative spacetime -- but this is not what happens.  Instead,
the non-associativity merely manifests itself by permitting new gauge
groups and fermionic representations to be obtained from the spectral
action.  Ultimately, as we shall explain, this is because the symmetry
group which appears in the spectral action is the automorphism group
of the underlying algebra, and this automorphism group is always an
ordinary associative group, even when the algebra itself is
non-associative.  The bosonic fields which arise in the
spectral action come from the requirement that the formalism should be
covariant with respect to the automorphisms of the underlying algebra, and
so they are built from the corresponding derivation operators (the 
infinitessimal generators of those automorphisms) in a way
that continues to be perfectly sensible and unambiguous, even when the
underlying algebra is non-associative.  In section \ref{sec4}, for
illustration, we present the simplest almost-associative geometry,
based on the octonions, and work out the spectral action in this case
as a proof of principle.

In future works, \cite{FarnsworthBoyle, Boyle:2014wba,
  Farnsworth:2014vva} we take these ideas much further. In 
  Ref.~\cite{FarnsworthBoyle}, we show how to construct non-associative
geometries corresponding to more realistic physical models that include 
Higgs fields, spontaneous symmetry breaking and fermion masses.  In 
another follow up paper \cite{Boyle:2014wba}, we show how a development
of the formalism initiated here can be applied to Connes and Chamseddine's construction of the
standard model, where it leads to a unification and simplification of many of the traditional
NCG axioms, together with a new geometric constraint on the finite-dimensional part of the 
Dirac operator which resolves a long-standing problem with that (otherwise strikingly 
successful) construction. Then, in Ref.~\cite{Farnsworth:2014vva}, we show that the same 
formalism suggests that the standard model of particle physics should be extended by two 
new particles -- a $U(1)_{B-L}$ gauge 
boson and a complex scalar field that carries charge $B-L=2$ and 
is responsible for "higgsing" the new $U(1)_{B-L}$ gauge symmetry;
this extension is experimentally viable, fixes the conflict between the observed 
Higgs mass and the value traditionally predicted by NCG, and also has other cosmological 
consequences that we are currently analyzing.

For earlier work on nonassociative geometry in different contexts, see
\cite{Wulkenhaar:1998sc, Wulkenhaar:1996pv, Wulkenhaar:1997jy,
  Wulkenhaar:1996av, 2004JMP....45.3883A, 2013MPLA...2850163G, 
  Blumenhagen:2011ph, Plauschinn:2012kd, Lust:2012fp, Blumenhagen:2013zpa}.

\section{Preliminaries}
\label{MathPrelims}

The purpose of this paper is to extend the formalism of non-commutative geometry.  First, however,
we must briefly review the elements of the associative NCG formalism, to discuss which elements need to be reformulated and generalized.  In sub-section \ref{spectral_triple} we give a very brief introduction to spectral triples and a short overview of the associative NCG formalism (for more details, see \cite{vandenDungen:2012ky}). We introduce two example spectral triples that will be important later in the paper. In sub-section \ref{Definitions}, we briefly introduce non-associative $\ast$-algebras, along with their automorphisms and derivations; and we meet the non-associative $\ast$-algebra that will serve as our main example in this paper: $\mathbb{O}$, the algebra of octonions. Although there is already a rich literature on both topics (see {\it e.g.}\ \cite{ConwaySmith, 2001math......5155B,vandenDungen:2012ky,RSchaf} and references therein) this brief review draws together in one place those elements most necessary for understanding the remainder of the paper. In addition, the generalization we motivate requires a certain shift in perspective from the traditional approach to NCG, which is accompanied by a shift in notation, which we also outline here. 

\subsection{Spectral triples }
\label{spectral_triple}
To specify a geometry, one starts by giving some input data. In Riemannian geometry, the input data is the manifold ${\cal M}$ and its metric $g_{\mu\nu}$.  In NCG, one instead provides the corresponding input data in terms of a so-called `spectral triple' of elements:
\begin{align}
\{A,H,D\}.
\end{align}
Here $A$ is a $\ast$-algebra that is linearly represented on the Hilbert space $H$, while $D$ is another Hermitian operator on $H$.  Roughly speaking, one can think of $A$ and $D$ as carrying the (differential) topological and metric information, respectively.  A spectral triple may also be equipped with two additional operators $J$ and $\gamma$, which provide some additional structure to the geometry.  One is able to describe the dynamics of an NCG using the so called `spectral action' formula, which assigns a real number to a spectral triple \cite{Chamseddine:1996zu} (much as the Einstein-Hilbert action assigns a real number to an ordinary Riemannian geometry).

In order to form a valid spectral triple, the five elements $\{A,H,D\}$, $J$, and $\gamma$ must not be chosen arbitrarily.  Instead, they must satisfy certain axioms and assumptions that give the spectral triple its structure. The structure of a spectral triple may be described by building it up in five steps, adding one element at a time, and starting (i) by choosing a $\ast$-algebra $A$.  (ii) In Step 2, one chooses a (left) representation of $A$ on the Hilbert space $H$: the
representation is a map $\pi$ which takes each element $a\in A$ to a corresponding linear operator $\tilde{a}\equiv\pi(a)$ that acts on $H$.  The map $\pi$ must preserve the structure of $A$: {\it i.e.}\ it must be linear and satisfy $\pi(ab)=\pi(a)\pi(b)$ and $\pi(a^{\ast})=(\pi(a))^{\ast}$.  In the next three steps, we add three operators on $H$ ($D$, $\gamma$ and
$J$) and describe the constraints they must satisfy.  (iii) In Step 3, we add the hermitian
operator $D$: the commutator $[D,\pi(a)]$ must be bounded ($\forall
a\in A$).  (iv) In Step 4, we add the hermitian and unitary
operator $\gamma$: it must satisfy $\{\gamma,D\}=0$ and
$[\gamma,\pi(a)]=0$ ($\forall a\in A$).  (v) In Step 5, we add
the anti-unitary operator $J$: it satisfies $J^{2}=\epsilon$,
$JD=\epsilon' DJ$, and $J\gamma=\epsilon'' \gamma J$, where
$\epsilon$, $\epsilon'$ and $\epsilon''$ are three $\pm$
signs that depend on the so-called ``KO-dimension'' of the spectral
triple\footnote{See Definition 16 in section 6.8 of
  \cite{springerlinklandi}, or section 2.2.2 in
  \cite{vandenDungen:2012ky}.}.  
  
Given any operator $\tilde{a}=\pi(a)$, we can use $J$ to define a dual operator
$\tilde{a}^{0}=J\tilde{a}^{\ast} J^{\ast}$.  The interpretation is
that any $a\in A$ is represented in two ways: as an
operator $\tilde{a}$ that acts on $H$ from the left, and as an
operator $\tilde{a}^{0}$ that acts on $H$ from the right.  In
the traditional case, where $A$ is an {\it associative}
$\ast$-algebra, one then usually imposes two additional requirements given by the so called order conditions:
\begin{subequations}
  \label{order_zero_one_conditions}
  \begin{eqnarray}
    \label{order_zero_condition}
    \big[\tilde{a},\tilde{c}^{0}\big]&=&0\quad\forall\{a,c\}\in A
    \qquad\textrm{``the order zero condition,''} \\
    \label{order_one_condition}
    \big[\big[D,\tilde{a}\big],\tilde{c}^{0}\big]&=&0\quad\forall\{a,c\}\in A
    \qquad\textrm{``the order one condition.''}
  \end{eqnarray}
\end{subequations}
These last two axioms must be modified when the underlying
algebra $A$ is non-associative as we will discuss in subsection \ref{realstrucsec}. 

Next let us consider two examples of spectral triples satisfying the above axioms.  Both of these examples will play an important role when we describe the construction of almost-associative geometries later in the paper.

{\bf Example 1.}  The first example is the "canonical" spectral triple  $T_c = \{\mathcal{A}_{c},
\mathcal{H}_{c}, D_{c},\gamma_{c},J_{c}\}$.  Just as Riemannian geometry contains Euclidean geometry, 
non-commutative geometry contains Riemannian geometry, and reduces to Riemannian geometry for a special class of
spectral data: namely for the canonical spectral triples. The idea is that the Riemannian data $\{{\cal 
  M},g_{\mu\nu}\}$ and the canonical triple $T_c$ provide dual descriptions of the
same geometry, so that the canonical spectral triple may be obtained
from the Riemannian data, or vice versa. Starting from the Riemannian data $\{{\cal M},g_{\mu\nu}\}$, the
corresponding canonical spectral triple $T_c$ may be constructed as
follows: $A_{c}=C_{\infty}({\cal M})$ is the algebra of smooth
complex-valued functions on ${\cal M}$; $H_{c}=L^{2}({\cal
  M},S)$ is the Hilbert space of (square integrable) Dirac spinors on
$\{{\cal M},g_{\mu\nu}\}$; $D_{c}=\slash\!\!\!\!D=-i\gamma^{\mu} \nabla_{\mu}^{S}$ is
the ordinary curved-space Dirac operator on $\{{\cal M},
g_{\mu\nu}\}$; $\gamma_{c}$ is the helicity operator on $H_{c}$
({\it i.e.}\ what physicists usually call $\gamma_{5}$ in 4
dimensions); and $J_{c}$ is the charge conjugation operator on $H_{c}$.  As for the (left)
representation of $A_{c}$ on $H_{c}$, the functions $f\in A_{c}$ act on the spinor fields
$\psi\in H_{c}$ by pointwise multiplication: $\psi(x)\to f(x)\psi(x)$.  See 
\cite{vandenDungen:2012ky} for more details.

{\bf Example 2.} The second example is a finite dimensional,
non-commutative, associative spectral triple (as described e.g. in
\cite{Chamseddine:1996zu, vandenDungen:2012ky}) $T_F =\{A_F,H_F,D_F,
\gamma_F,J_F\}$. Here we take $A_{F}$ to be the algebra of $n\times n$ 
complex matrices $A_F = M_n(\mathbb{C})$ represented on 
themselves $H_F = M_n(\mathbb{C})$ in the natural way. The real structure 
element is the natural anti-linear involution on the input algebra 
({\it i.e.}\ hermitian conjugation) $J_{F}h = h^{\dagger}\;(h\in H_{F}=M_{n}(\mathbb{C}))$, and the grading 
operator is the $n\times n$ identity matrix $\gamma_F = \mathbb{I}$. 
The condition $\{\gamma_F,D_F\} = 0$ then implies that the 
$n\times n$ hermitian matrix $D_{F}$ is equal to zero. 

Finally, given any two spectral triples, $T_{1}=\{A_{1},H_{1},D_{1}\}$ and
$T_{2}=\{A_{2},H_{2},D_{2} \}$, we can construct a third spectral triple
$T_{12}=\{A_{12},H_{12},D_{12}\}$, where $A_{12}=A_{1}\otimes A_{2}$, $H_{12}=H_{1}\otimes H_{2}$ and $D_{12}=D_{1}\otimes \mathbb{I}_{2}+\gamma_{1}\otimes D_{2}$.\footnote{Strictly speaking, this is the formula for the product of two {\it even} spectral triples: for the more general formula, see \cite{2011IJGMM..08.1833D}.}  For so called  `even' and `real' spectral triples, which are additionally equipped with the operators $\gamma$ and $J$, one also defines
$\gamma_{12}=\gamma_{1}\otimes\gamma_{2}$ and $J_{12}=J_{1}\otimes J_{2}$. 
The product $T_{c}\times T_{F}$ of an infinite-dimensional canonical spectral triple $T_c$ with a
finite-dimensional non-commutative spectral triple $T_F$ is called an "almost commutative geometry". 
For the almost commutative geometry where $T_{F}$ is the simple example geometry 
given above, the corresponding spectral action is Einstein gravity coupled to $SU(N)$
Yang-Mills theory \cite{Chamseddine:1996zu}.  Einstein gravity coupled to the full 
standard model of particle physics comes from evaluating the spectral action for an almost commutative 
geometry involving a slightly more complicated finite geometry $T_{F}$ \cite{vandenDungen:2012ky}.  

When the finite dimensional algebra $A_{F}$ is non-associative, we
will refer to $T_c\times T_F$ as an ``almost-associative geometry.''  In Section \ref{sec4} we 
will present the simplest example of an almost-associative geometry (based on the algebra 
$A_{F}=\mathbb{O}$) and give its spectral action.

\subsection{$\ast$-algebras, automorphisms, and derivations}
\label{Definitions}

When constructing an NCG, the primary input is an algebra $A$. In this subsection we will briefly introduce algebras, and introduce notation for the remainder of the paper, paying particular attention to the general non-associative case.

An algebra $A$ is a vector space over a field $\mathbb{K}$,
which is equipped with an additional binary ``product'' operation: a
$\mathbb{K}$-bilinear map from $A\times A\to A$.
(The product of two vectors $a,b\in A$ is another vector
$ab\in A$.)  A $\ast$-algebra $A$ is an algebra that is
equipped with an additional anti-linear involution map
$*:A\rightarrow A$ satisfying
\begin{equation}
  (a^{\ast})^{\ast}=a,\qquad (ab)^* = b^*a^*,
  \qquad a,b\in A.
\end{equation}
In a $\ast$-algebra $A$, the elements satisfying $u^{\ast}=u^{-1}$, $h^{\ast}=h$ or
$a^{\ast}=-a$ are called "unitary," "hermitian," or "anti-hermitian," respectively.

When we say an algebra is non-commutative, we mean its product is
non-commutative: $ab\neq ba$, $a,b\in A$.  Similarly, when we
say an algebra is non-associative, we mean that its product is
non-associative: $(ab)c\neq a(bc)$, $a,b,c\in A$.  Just as we
introduce the ``commutator'' $[a,b]$ to characterize the failure of
commutativity, we introduce the ``associator'' $[a,b,c]$ to
characterize the failure of associativity
\begin{align}
  [a,b]\equiv ab-ba,\qquad
  [a,b,c]\equiv (ab)c - a(bc)
  \qquad a,b,c\in A.
\end{align}

Lie algebras are familiar examples of non-associative algebras.  For
example, consider the vector space of $N\times N$ complex
anti-hermitian matrices.  These do not form an algebra under ordinary
matrix multiplication (since the ordinary matrix product of two
anti-hermitian matrices is not, in general, anti-hermitian), but they
do form an algebra if we define the "product" $[a,b]$ to be the commutator
of the matrices $a$ and $b$.  The resulting algebra
is a Lie algebra, since the product $[a,b]$ is anti-symmetric and
satisfies the Jacobi identity $[[a,b],c]+[[b,c],a]+[[c,a],b]=0$.  But it is easy to
check that this product is non-associative: $[[a,b],c]\neq [a,[b,c]]$.

It will be convenient to introduce the standard notation given in \cite{RSchaf} in 
which $L_{a}$ denotes the left-action of $a$, and $R_{a}$ denotes the right-action of $a$:
\begin{equation}
  L_{a}b\equiv ab\qquad R_{a}b\equiv ba,\qquad a,b\in A.\label{leftright}
\end{equation}
In other words, $L_{a}$ and $R_{a}$ are two different linear operators
on the vector space $A$.  As an illustration of this notation
we can write $a((cv)b)=L_{a}R_{b}L_{c}v$ (with $a,b,c,v\in A$).
In particular, note that when $A$ is nonassociative, the
left-hand side of this equation requires parentheses, but the
right-hand side does not.  For our later discussions, it is important
to keep in mind that $L_{a}$ and $R_{a}$ are associative operators, 
even when $A$ is non-associative.  Also note that in the 
non-associative case $L_a L_b\neq L_{ab}$. 

If $A$ is a $\ast$-algebra, then an {\it automorphism} of
$A$ is an invertible linear map $\alpha:A\to A$
which respects the product and involution operations in $A$:
\begin{equation}
  \alpha(ab)=\alpha(a)\alpha(b)\qquad\alpha(a^{\ast})=
  (\alpha(a))^{\ast},
\end{equation}
and a {\it derivation} of $A$ is a linear map $\delta:A
\to A$ which satisfies
\begin{equation}
  \delta(ab)=\delta(a)b+a\delta(b)\qquad\delta(a^{\ast})=
  (\delta(a))^{\ast}.
\end{equation}
Note that, when the automorphism $\alpha$ is infinitessimally close to
the identity map ``${\rm Id}$,'' it can be written as $\alpha={\rm
  Id}+\delta$ where $\delta$ is a derivation.  The derivations of
$A$ are the infinitessimal generators of the automorphisms of
$A$; they form a Lie algebra, with Lie bracket given by
$[\delta_{1},\delta_{2}]=\delta_{1}\circ\delta_{2}-
\delta_{2}\circ\delta_{1}$ (where $\circ$ denotes composition of
operators).

As a first example, consider the $\ast$-algebra which appears in the canonical 
spectral triple $T_{c}$ introduced in section \ref{spectral_triple} -- {\it i.e.}\ the
$\ast$-algebra $A=C_{\infty}({\cal M},\mathbb{C})$ 
of smooth functions from a manifold ${\cal M}$ to the
complex numbers $\mathbb{C}$. In this case the automorphisms
$\alpha_{\varphi}:A\to A$ are nothing but the maps
$\alpha_{\varphi}(f)=f\circ\varphi$, where $f:{\cal M}\to\mathbb{C}$
is a smooth function and $\varphi:{\cal M}\to{\cal M}$ is a
diffeomorphism; and if we consider the automorphisms
infinitessimally close to the identity, we see that the corresponding
derivations have the form $\delta_{v}(f)=v^{\mu}\partial_{\mu}f$,
where $v^{\mu}(x)$ is a contravariant vector field on ${\cal M}$.

Next consider a more general associative (but possibly non-commutative) 
$\ast$-algebra $A$.  Within its full group  of automorphisms, ${\rm Aut}(A)$, there
is a normal subgroup ${\rm Inn}(A)$ of ``inner automorphisms''
of the form $\alpha_{u}(a)=u a u^{\ast}$, where $u\in A$ is unitary.  The
group of ``outer automorphisms'' is then defined to be the quotient
${\rm Out}(A)={\rm Aut}(A)/{\rm Inn}(A)$.
If we note that the unitary elements $u$ are generated by
anti-hermitian elements $a\in A$ ($u={\rm e}^{a}$) and study the inner
automorphisms infinitessimally close to identity map, we find that the corresponding 
`inner' derivations ({\it i.e.}\ the generators of the inner
automorphisms) are $\delta_{a}(b)=[a,b]$ or equivalently:
\begin{equation}
  \delta_{a}=L_{a}-R_{a}.\label{assocder}
\end{equation}

In particular, in the spectral reformulation of Einstein gravity coupled to the
standard model of particle physics, the idea is (roughly) the
following: one starts from a $\ast$-algebra $A=C_{\infty}({\cal M},
A_{F})$, whose full automorphism group ${\rm Aut}(A)$ corresponds to 
the full (gauge+gravitational) symmetry group of the spectral action; roughly
speaking, the inner automorphisms ${\rm Inn}({\cal A})$ are the group
of maps from ${\cal M}$ to ${\rm Aut}(A_{F})$ and correspond to
the group of gauge transformations, while the outer automorphisms
${\rm Out}(A)={\rm Diff}({\cal M})$ correspond to the group of gravitational
symmetries ({\it i.e.}\ the diffeomorphisms of ${\cal M}$)).  In the subsequent
sections, we will explain that a very similar story obtains when the 
finite-dimensional algebra $A_{F}$ is non-associative.

Finally just as non-commutative associative algebras have inner automorphisms, so too in general do non-associative algebras. In this paper, for the purposes of illustration, we will focus on one
of the most famous finite-dimensional nonassociative algebras: namely, the algebra
$\mathbb{O}$ of octonions.  The octonions occupy a special place in
mathematics.  They are one of only four normed division algebras (along side
the real numbers $\mathbb{R}$, the complex numbers $\mathbb{C}$ and the
quaternions $\mathbb{H}$).  The algebras $\mathbb{R}$, $\mathbb{C}$, 
$\mathbb{H}$, and $\mathbb{O}$ are, respectively, 1, 2, 4 and 8 dimensional, 
with 0, 1, 3 and 7 imaginary elements which square to negative one.  The octonions 
are intimately connected to some of the most beautiful structures in mathematics,
including the exceptional Lie algebras and the exceptional Jordan
algebra.  For nice introductions to the octonions, and
their connections to other areas of mathematics, see
\cite{ConwaySmith, 2001math......5155B}.  Here, let us note three features in
particular.  (i) First, the octonions are an example of an
``alternative algebra'' -- {\it i.e.}\ an algebra in which the
associator $[a,b,c]$ flips sign under interchange of any two of its
arguments.  (ii) Second, the general derivation $\delta:A\to A$ may be written 
as a linear combination of derivations of the form \cite{RSchaf}
\begin{equation}
  \label{fmnonassocder}
  \delta_{a,b}=[L_{a},L_{b}]+[L_{a},R_{b}]+[R_{a},R_{b}]\qquad
  a,b\in A.
\end{equation}
[In fact, the general inner derivation of any alternative algebra may be written this way.  
Again, it is important to emphasize that although the alternative algebras are in general non-associative, 
their derivations {\it are} associative (as is the case for all algebras); and in particular the inner derivations $\delta_{a,b}$ are constructed from the associative operators  $L_{a}$ and $R_{a}$ defined in \eqref{leftright}. 
One can check that these derivations form a Lie algebra under 
the Lie bracket, and generate an associative Lie group under exponentiation.]  (iii) Third, the 
algebra of derivations of $\mathbb{O}$ is $g_{2}$ (the smallest exceptional Lie algebra) and the 
automorphism group of $\mathbb{O}$ is $G_{2}$ (the smallest exceptional Lie group).

\section{Non-Associative Geometry}
\label{NAGformalism}

Now we discuss how the structure of a spectral triple must be generalized in the 
case where the algebra $A$ is non-associative.  The organization of this section is as follows:

(i) In Subsection \ref{representing}, we clarify what it means to represent a non-associative $\ast$-algebra
$A$ on a Hilbert space $H$.

(ii) In Subsection \ref{autocov}, we articulate the principle of $\ast$-automorphism covariance, which ties together the transformations of the input algebra $A$ with those of the Hilbert space $H$, and all of the operators that act on it. The principle of $\ast$-automorphism covariance subsumes and replaces the traditional covariance principles of physics: diffeomorphism covariance (in Einstein gravity) and gauge covariance (in gauge theory).

(iii) In Subsection \ref{realstrucsec}, we re-introduce the grading and real structure operators $\gamma$ and $J$ in the non-associative setting.  In particular we explain how the usual ``order zero''  condition given in equation \ref{order_zero_one_conditions} generalizes in the non-associative case. We address the higher order conditions in future papers \cite{FarnsworthBoyle, Boyle:2014wba,
  Farnsworth:2014vva}.
  
(iv) In Subsection \ref{flucDir}, we explain how to obtain a `fluctuated' Dirac operator $D_A$ from an `un-fluctuated' Dirac operator $D$.  Just as one creates a covariant derivative in regular gauge theory by adding a one-form built from the
generators of the underlying symmetry group, in spectral geometry, one creates a covariant Dirac operator by adding a `one-form' built from the derivations of the underlying $\ast$-algebra. 

\subsection{Representing a non-associative $\ast$-algebra}
\label{representing}

The starting point for the spectral formalism is a $\ast$-algebra $A$ that is represented (or, more correctly,
`bi-represented' -- {\it i.e.}\ represented from both the left and the right) on $H$.  But, in attempting to 
extend the definition of a bi-representation of $A$ on $H$ to the case where $A$ is non-associative, we 
seem to encounter a problem.  
After all, by a representation of $A$ on $H$ we usually mean a linear map from each element 
$a\in A$ to a linear operator $\pi(a)$ on $H$, such that the composition of such operators represents the 
product on $A$: $\pi(a)\pi(b)=\pi(ab)$.  Yet the composition of linear operators is associative, so it seems 
that we cannot possibly represent the non-associativity of $A$ in this way.

The elegant solution to this problem (originally due to Samuel Eilenberg, we think, and nicely explained in 
Ch. II.4 of \cite{RSchaf}) involves a change of perspective.  For $a\in A$ and $h\in H$, we let $ah$
denote the left-action of $A$ on $H$ (a bilinear map from $A\times H\to H$); and similarly we 
let $ha$ denote the right-action of $A$ on $H$ (a bilinear map from $H\times A\to H$).  Now, 
given a class $\mathcal{C}$ of (possibly non-associative) algebras defined by a set of multi-linear 
identities $I_i(a_1,...,a_{n_{i}})=0$, and an algebra $A$ in $\mathcal{C}$, we say that $A$ is 
bi-represented on $H$ in $\mathcal{C}$ (or, equivalently, that $H$ is a bimodule over $A$
in $\mathcal{C}$) if all of the identities obtained by replacing any single $a_j \in A$ by any 
$h\in H$ are satisfied \cite{RSchaf}.  (For an alternative way of describing this generalization of 
bimodules to the non-associative case, see Ref.~\cite{Boyle:2014wba}.) 

As a first example, consider the class $\mathcal{C}$ of associative algebras.  An algebra $A$
in this class satisfies the multilinear identity
\begin{align}
[a_1,a_2,a_3] = 0 \label{assoc}\quad(\forall a_{i}\in A).
\end{align}  
Replacing any one algebra element in \eqref{assoc} with a vector space element $h\in H$, we obtain following conditions
\begin{subequations}
  \label{assrls}
  \begin{align}
  \label{aah}
  [a_1,a_2,h] &=0, \\
  \label{haa}
  [h,a_{2},a_{3}] &=0,\\
  \label{aha}
  [a_1,h,a_3] &=0,
  \end{align}
\end{subequations}
for all $a_i\in A$ in $h\in H$.  Here $ah\in H$ denotes the left action of $A$ on $H$, while $ha\in H$ denotes
the right action of $A$ on $H$.  Note that Eq.~(\ref{aah}) is just an unfamiliar way of phrasing the familiar
fact that $A$ is left-represented on $H$: $\pi(ab)=\pi(a)\pi(b)$.  Similarly, Eq.~(\ref{haa}) says that
$A$ is right-represented on $H$.  Finally, Eq.~(\ref{aha}) says that the left and right representations
of $A$ on $H$ commute with one another, in the sense of the order zero condition (\ref{order_zero_condition}).
Thus, in this case we see that these three conditions, together, simply recover the usual definition of an associative 
bi-representation of $A$ on $H$.

Following equations \eqref{assrls}, the products between elements in an associative representation $\tilde{a},\tilde{b}\in \pi(A)$ will be given by composition $\tilde{a}\tilde{b} = \tilde{a}\circ\tilde{b}$. Composition is associative, and so expressions like $\tilde{a}\tilde{b}\tilde{c}$ and
$\tilde{a}\tilde{b}h$ are unambiguous, and do not require any
additional parentheses.  By contrast, in the case where $A$ is non-associative, the operator $\tilde{a}$ has two different roles that should be carefully distinguished: on the one hand it can operate on a vector $h\in H$, mapping it to a new vector $\tilde{a}h\in H$; on the other hand, it can multiply another operator $\tilde{b}$
to form a third operator $(\tilde{a}\tilde{b})$.  It is important to
note that, since the operators $\tilde{a}$ and $\tilde{b}$ represent
elements in an underlying non-associative algebra $A$, their
product $(\tilde{a}\tilde{b})$ will {\it not} be given by the composition
of the operators $\tilde{a}$ and $\tilde{b}$  on $H$ (which is associative);
instead, it will be given by some other product that reflects the
non-associativity of $A$. In left-right notation, this is again the statement that $L_{\tilde{a}}L_{\tilde{b}}\neq L_{\tilde{ab}}$ for non-associative input algebras $A$. 

Finally, the vector spaces $H$ that we will be dealing with will also be Hilbert spaces in the sense that they will be equipped with an inner product $\langle\;|\;\rangle$ -- a rule for
multiplying two vectors $a$ and $c$ to get a scalar $\langle
a|c\rangle\in\mathbb{K}$.  The inner product is skew-linear in its
first argument, linear in its second argument, skew-symmetric
($\langle a|c\rangle=\langle c|a\rangle^{\ast}$), and positive
definite ($\langle a|a\rangle\geq0$). As a simple illustration, consider the case where $A$ is a
non-associative $\ast$-algebra equipped with a natural inner product
$\langle\;|\;\rangle$ (so that it may also be interpreted as a Hilbert
space $H$).  Then $A$ may be `represented on itself'
in an obvious way: we take the Hilbert space $H$ to be the same
as the $\ast$-algebra $A$; we take the algebra homomorphism $\pi$ to be the identity map ($\tilde{a}=a$); and we take the product of two operators $\tilde{a}$ and $\tilde{b}$, or the action of an operator $\tilde{a}$ on a Hilbert space element $h$, to be given by the underlying product in $A$: $\tilde{a}\tilde{b}=ab$, $\tilde{a}h=ah$.  In the example finite non-associative geometry we consider in Section \ref{sec4}, we take $A=H=\mathbb{O}$ the algebra of octonions, which is equipped with a natural inner product $\langle a|b\rangle=(1/2)(a^{\ast}b+b^{\ast}a)= {\rm Re}(a^{\ast}b)$ where $a^{\ast}$ is the octonionic conjugate of $a$.

\subsection{The principle of automorphism covariance}
\label{autocov}

Consider an automorphism $\alpha$ of the input $\ast$-algebra $A$,
which maps each element $a\in A$ to a new element $a'\in A$.  This induces a corresponding transformation $\tilde{\alpha}$
that maps each operator $\tilde{a}$ to a new operator $\tilde{a}'$, and a corresponding transformation $\hat{\alpha}$ that maps each vector $h\in H$ to a new vector $h'\in h$:
\begin{subequations}
  \begin{eqnarray}
    a&\to&a'=\alpha(a), \\
    \tilde{a}&\to&\tilde{a}'=\tilde{\alpha}(\tilde{a}), \\
    h&\to&h'=\hat{\alpha}(h).
  \end{eqnarray}
\end{subequations}
To tie the transformations $\alpha$, $\tilde{\alpha}$ and
$\hat{\alpha}$ together, we demand that they satisfy the {\it
  principle of automorphism covarariance}, which demands that our
whole formalism should ``commute'' with automorphisms of the
underlying $\ast$-algebra.  In other words, any sensible expression
should have the property that if we first transform its components
and then evaluate the expression, this should be the same as first
evaluating the expression and then transforming the result.

For starters, we apply the principle to the expression $\tilde{a}
=\pi(a)$: it requires that $\pi(\alpha(a))=\tilde{\alpha} (\pi(a))$,
$\forall a\in A$; or, in other words:
\begin{equation}
  \pi\circ\alpha=\tilde{\alpha}\circ\pi
\end{equation}
where $\circ$ denotes composition of functions.  Next, we apply the
principle to the expression $\tilde{a}h$: it requires that
$\hat{\alpha}(\tilde{a}h)= \tilde{\alpha}(\tilde{a})\hat{\alpha}(h)$;
or, in other words:
\begin{equation}
  \tilde{a}'=\tilde{\alpha}(\tilde{a})
  =\hat{\alpha}\circ\tilde{a}\circ\hat{\alpha}^{-1}
  \qquad\forall a\in A.
\end{equation}

For illustration, consider the simple example of an algebra represented on itself $H = A$.  In this case, we would have
$\alpha=\tilde{\alpha}=\hat{\alpha}$, and all of the above equations
would be automatically satisfied.

When constructing a spectral geometry, there are three other important operators which act
on $H$: namely, $D$, $\gamma$, and $J$.  Applying the principle
to the expressions $D h$, $\gamma h$ and $J h$ we see that, under an
automorphism $\alpha$, these operators must transform as
\begin{subequations}
  \label{Trans}
  \begin{eqnarray}
    \label{TransD}
    D&\to&\!D'=\hat{\alpha}\circ\!D\circ \hat{\alpha}^{-1}, \\
    \label{TransG}
    \gamma&\to&\gamma'\;\!=\hat{\alpha}\circ\gamma
    \;\!\circ\hat{\alpha}^{-1}, \\
    \label{TransJ}
    J&\to&J'=\hat{\alpha}\circ J\circ\hat{\alpha}^{-1}.
  \end{eqnarray}
\end{subequations}
In fact, as we shall see, $J$ and $\gamma$ are naturally invariant
under this transformation ({\it i.e.}\ $J'=J$ and $\gamma'=\gamma$);
but $D$ is not: instead, the automorphisms of the underlying
$\ast$-algebra $A$ induce a transformation or ``fluctuation''
of $D$, from which the bosonic fields arise.

We take the principle of automorphism covariance to be a fundamental
principle lying at the base of the spectral reformulation of physics:
as we shall see, it replaces (or subsumes or implies) the more
familiar principles of covariance under coordinate transformations and
gauge transformations, which are usually taken as the starting points
for Einstein gravity and gauge theory.  We will also see that this
principle will give us all the guidance we need in formulating the
spectral action principle unambiguously, even when $A$ is
nonassociative.

\subsection{The real structure $J$, and the $\mathbb{Z}_{2}$ grading
  $\gamma$}
\label{realstrucsec}
A spectral triple is said to be ``real'' if it is equipped with a real
structure operator $J$ and ``even'' if it is equipped with a
$\mathbb{Z}_{2}$ grading operator $\gamma$. In this section we will
discuss the generalization of both operators to the non-associative
setting, beginning with the operator $J$. For a more complete
exposition in the associative case see references
\cite{Connes:1995tu,Connes:1996gi, vandenDungen:2012ky}.

First consider the real structure $J$.  The basic observation, which
remains perfectly valid when $A$ is non-associative, is that we can think
of $J$ as extending the $\ast$ operation from the $\ast$-algebra $A$ to 
the bimodule $H$ over $A$.  So we introduce the anti-linear operator
$J$ on $H$ to define the transformation $(h\in H) \to (Jh\in H)$, 
which parallels the anti-linear operation $(a\in A)\to(a^{\ast}\in A)$.  
And, since $\ast$ is an anti-automorphism on $A$, so that it 
acts on any product of algebra elements $a,b\in A$ as  $(ab)^{\ast}=
b^{\ast}a^{\ast}$, $J$ must have a compatible action on any product
of algebra elements $a\in A$ and a Hilbert space elements $h\in H$:
in particular, $J(ah)=(Jh)a^{\ast}$ and $J(ha)=a^{\ast}(Jh)$.  In other words, we recover
Connes' familiar relations between left action and right action
\begin{subequations}
  \begin{eqnarray}
    R_{a}&=&J L_{a^{\ast}}J^{\ast} \\
    L_{a}&=&J R_{a^{\ast}}J^{\ast}
  \end{eqnarray}
\end{subequations}
and see that they remain unchanged in the non-associative case. 

$J$ also plays an important role in Connes' so-called order-zero and order-one 
conditions \eqref{order_zero_one_conditions}.  But, from the perspective presented here, 
we can see that these are really assumptions about the associativity of the bimodule 
$H$ over $A$,\footnote{or, more generally, the bimodule $H$ over $\Omega A$, where 
$\Omega A$ is the differential graded $\ast$-algebra of forms over $A$: this 
generalization is treated in detail in our subsequent papers \cite{FarnsworthBoyle, 
Boyle:2014wba, Farnsworth:2014vva}} which must be appropriately modified 
in the case where $A$ is non-associative.  In particular, note that, from our current
perspective, the traditional order-zero condition \eqref{order_zero_condition} 
directly follows from the traditional assumption \eqref{assrls} that $H$ is an 
{\it associative} bimodule over $A$ -- in particular, it is nothing but the assumption
that the associator $[a_{1},h,a_{2}]$ must vanish for any $a_{1},a_{2}\in A$ and 
$h\in H$. 

This traditional order-zero condition \eqref{order_zero_condition} is no longer 
appropriate in the case  when $A$ is non-associative.  To clarify this point,
consider as an example the case where the input data is $A=H=\mathbb{O}$,
with the octonions acting on themselves in the obvious way and $J$ is just
octonionic conjugation.  In this case we find
\begin{align}
[JL_{\tilde{b}^*}J^*,L_{\tilde{a}}]\tilde{h} = [R_{\tilde{b}},L_{\tilde{a}}]\tilde{h} = 
[\tilde{a},\tilde{h},\tilde{b}]\neq 0,\hspace{1cm}a,b\in A,
\hspace{2mm}h\in H \label{associator}
\end{align}
As the octonions are non-associative, the associator is typically non-zero, 
so we see that the traditional order-zero condition \eqref{order_zero_condition} 
is incompatible with the representation of the the octonions on themselves, 
which is the most natural representation.  A similar `problem' will arise for all 
non-associative algebras and their representations. 

Fortunately, subsection \ref{representing} points to the solution: although a non-associative algebra represented on a Hilbert space may not satisfy the traditional (associative) order zero condition, it will instead satisfy a set of conditions appropriate to the associativity class to which it belongs. The bimodule given above $A = H = \mathbb{O}$ will for example satisfy the alternative order zero conditions, because the octonions are an alternative algebra:
\begin{subequations}
\begin{align}
[R_{\tilde{b}},L_{\tilde{a}}] &= [L_{\tilde{b}},R_{\tilde{a}}],\label{altor01}\\
[R_{\tilde{b}},L_{\tilde{a}}] &= L_{\tilde{b}\tilde{a}} - L_{\tilde{b}}L_{\tilde{a}} =  R_{\tilde{a}}R_{\tilde{b}}-R_{\tilde{a}\tilde{b}}. \label{altor02}
\end{align}
\end{subequations}
The main purpose of the order-$n$ conditions is to ensure automorphism covariance. In an associative NCG, the associative order zero condition ensures that the inner derivations may always be written without loss of generality in the associative form given in \eqref{assocder}.  More generally, the
order zero condition, and higher order conditions, along with the operator $J$ define the bi-module  structure of the Hilbert space $H$, and ensure covariance under the automorphisms of the input algebra regardless of its associativity properties. In this paper we will not have any use for the order one condition (or higher order conditions) and so we will leave their elucidation to future work \cite{FarnsworthBoyle, Boyle:2014wba,
  Farnsworth:2014vva}.

Now let us turn to the $\mathbb{Z}_{2}$ grading $\gamma$.  It is a linear
operator on $H$ that commutes with the action of $A$ on $H$.  It is both hermitian 
($\gamma^{\ast}=\gamma$) and unitary ($\gamma^{\ast}=\gamma^{-1}$):
hence it satisfies $\gamma^2=1$, so its eigenvalues are $\pm1$, and 
it correspondingly decomposes $H$ into two subspaces $H=H_{+}\oplus H_{-}$.
Note that all of these defining properties continue to make perfect sense when 
$A$ is non-associative, and require no modification.

For physicists, the familiar example is Dirac's helicity operator
$\gamma_{5}$ which has the above properties and decomposes the space
of Dirac spinors into positive and negative (helicity) subspaces:
$L^2({\cal M},S)=L^2_{+}({\cal M},S)+L^2_{-}({\cal M},S)$.
Another nice way to think of $\gamma_5$ is as a volume form. This 
perspective may also be generalized to the non-commutative
and non-associative cases. Recall that on a spin manifold the Dirac
operator is given by $\slashed{D} = -i\gamma^\mu\triangledown_\mu^S$,
where the $\gamma^\mu$ are the Dirac Gamma matrices, and
$\triangledown_\mu^S$ is the Levi-Civita connection on the spinor bundle. 
Although this Dirac operator may be unbounded, its commutator with elements of the algebra
of functions over the manifold $df = [\slashed{D},f] =
-i\gamma^\mu(\partial_\mu f)$ is bounded. In fact this bounded
operator gives the Clifford representation of the 1-form $df =
dx^\mu(\partial_\mu f)$ \cite{vandenDungen:2012ky}.  Similarly, we see
that the $\gamma_5$ grading operator in the canonical case can be
considered as the Clifford representation of a volume form.
\begin{align}
\tfrac{1}{4!}\epsilon_{\mu\nu\tau\rho}\gamma^\mu\gamma^\nu\gamma^\tau\gamma^\rho = 
\gamma^1\gamma^2\gamma^3\gamma^4 := \gamma_5.
\end{align}
Connes generalized this grading structure to non-commutative even
dimensional orientable spin manifolds \cite{Connes:1996gi}. Relatedly,
he introduced a new differential calculus which generalizes the De Rham 
cohomology of ordinary differential calculus to what is known as cyclic 
cohomology \cite{conNCG, conmarNCG}. When extending to the non-associative 
case further generalization is necessary.  Fortunately, much work has already 
been done in this direction. As a description of this generalization will not be 
necessary for understanding our first example non-associative geometry 
we will not give an account of it here, and instead refer to the interested reader 
to the literature \cite{2001math.....10199K, 2004JMP....45.3883A}.
 
Both the real structure $J$ and the $\mathbb{Z}_{2}$ grading $\gamma$
should be compatible with the automorphisms of the underlying
$\ast$-algebra: automorphisms should not affect the split between
positive and negative helicity states, or between particles and
anti-particles.  We can express this requirement in terms of
automorphisms:
\begin{subequations}
  \begin{eqnarray}
    \gamma'&=&\hat{\alpha}\circ \gamma\circ\hat{\alpha}^{-1}=\gamma, \\
    J'&=&\hat{\alpha}\circ J\circ\hat{\alpha}^{-1}=J.
  \end{eqnarray}
\end{subequations}
or in terms of the derivations that generate them
\begin{subequations}
  \begin{eqnarray}
    \big[ \tilde{\delta},\gamma\big] &=&0 \\
    \big[ \tilde{\delta}, J\big] &=&0
  \end{eqnarray}
\end{subequations}
Readers can convince themselves that these conditions do indeed hold for inner derivations in the
associative case, and in the nonassociative example that will be discussed below in section \ref{sec4}.
We propose that it is natural to take these conditions to be true
more generally; {\it i.e.}\ to take them as axiomatic in
non-associative geometry.

\subsection{Fluctuating the Dirac operator $D$}
\label{flucDir}

We are finally able to discuss the fluctuations of the Dirac operator corresponding to a non-associative geometry. In ordinary gauge theory, the principle of gauge covariance leads us
to replace the partial derivative $\partial_{\mu}$ by the gauge
covariant derivative $D_{\mu}=\partial_{\mu}+A_{\mu}$, which is
ultimately the object from which we build a gauge-invariant action.
In a closely analogous way, in spectral geometry the principle of
$\ast$-automorphism covariance leads us to replace the fiducial
``Dirac operator'' $D$ with the ``fluctuated'' or ``$\ast$-algebra
covariant'' Dirac operator $D_A$, which is ultimately the object
from which we build the the $\ast$-automorphism-invariant spectral
action.

It is helpful, then, to warm up by reviewing the story in ordinary
gauge theory.  We can write a general gauge transformation in the form
$u(x)={\rm exp}[\alpha^{a}(x)T_{a}]$, where $T_{a}$ are the generators
of the gauge group.  Now consider a multiplet of matter fields $\psi$
that transforms covariantly under a gauge transformation:
$\psi\to\psi'=u\psi$.  We would like to introduce a gauge-covariant
derivative operator $D_{\mu}$ with the property that $D_{\mu}\psi$
also transforms covariantly: $D_{\mu}\psi\to D_{\mu}'\psi'=
uD_{\mu}\psi$.  In other words, we want $D_{\mu}$ to transform as
\begin{equation}
  \label{D_mu_transform_exact}
  D_{\mu}\to D_{\mu}'=uD_{\mu}u^{-1}.
\end{equation}
Start with the special case where $D_{\mu}=\partial_{\mu}$, and
perform an infinitessimal gauge transformation to obtain
$D_{\mu}'=\partial_{\mu}-[\partial_{\mu},\alpha^{a}(x)]T_{a}$.  By
inspection of this formula, we see that in the general case we can
take
\begin{equation}
  D_{\mu}=\partial_{\mu}+B_{\mu}\qquad{\rm where}\qquad
  B_{\mu}=B_{\mu}^{a}T_{a}. \label{covdiv}
\end{equation}
Here $B_{\mu}^{a}$ are arbitrary gauge fields (one for each linearly
independent generator $T_{a}$). To make $D_{\mu}$ transform as in
Eq.~(\ref{D_mu_transform_exact}), we should take $B_{\mu}$ to
transform as
\begin{equation}
  B_{\mu}\to B_{\mu}'=uB_{\mu}u^{-1}+u[\partial_{\mu},u^{-1}].
\end{equation}

Now let us present the analogous story in spectral
geometry, in which inner automorphisms act to `fluctuate' the Dirac operator $D\rightarrow D_A$. To begin with, consider an element $h\in H$; under an inner
$\ast$-automorphism $\alpha$ of $A$ it transforms as $h\to
h'=\hat{\alpha}(h)$ (for the relationship between the hatted and un-hatted transformations, see subsection \ref{autocov}).  We would like to introduce a $\ast$-automorphism-covariant Dirac operator
$D_A$ such that $D_A h$ also transforms as $D_A h\to
D_A'h'=\hat{\alpha}(D_A h)$.  In other words,  just as in equation \eqref{TransD}, the covariant Dirac operator $D_A$ must transform as
\begin{equation}
  D_A\to D_A'=\hat{\alpha}\circ D_A \circ
  \hat{\alpha}^{-1}.\label{flucdirtran}
\end{equation}
In the case of almost-associative or almost-commutative geometries, the input algebra
$C^{\infty}({\cal M},A_{F})$ is the algebra of functions from the manifold ${\cal M}$ to the 
finite algebra $A_{F}$, and the inner automorphisms may be written as ${\alpha}=
{\rm exp}(\,{\delta}\,)$.  Here the corresponding inner derivations $\tilde{\delta}$ acting on $H$ 
may be written as  
$\tilde{\delta}=c^{i}(x)\otimes\tilde{\delta}_{i}$, where $\{\tilde{\delta}_{i}\}$ are a basis of derivations
for the finite algebra $A_{F}$, while $c^{i}(x)$ are spatially-varying coefficient functions
({\it i.e.}\ functions from ${\cal M}$ to $\mathbb{K}$, where $\mathbb{K}$ is the field over 
which $A_{F}$ is defined).  

If we apply \eqref{flucdirtran} to the Dirac operator for an almost-commutative or 
almost-associative geometry, $D=\slash\!\!\!\!D\otimes\mathbb{I}_{F}+\gamma_{c} 
\otimes D_{F}$ (see subsection \ref{spectral_triple}), and we 
expand $\hat{\alpha}={\rm exp}(\,\tilde{\delta}\,)$ to first order in the inner derivation 
$\tilde{\delta}$, we find
\begin{align}
  D'&\simeq D+[\tilde{\delta},D]=D
  -\underbrace{[\,\slash\!\!\!\!D,c^{i}(x)]\otimes\tilde{\delta}_{i}}_{\text{Gauge terms}}
  -\underbrace{\gamma_{c}\,c^{i}(x)\otimes[\tilde{\delta}_{i},D_{F}]}_{\text{Higgs terms}}.\label{GHsplit}
\end{align}
We see that in an almost-commutative or almost-associative geometry, we must
add both gauge fields {\it and also higgs fields} to the Dirac operator in order to make it
$\ast$-automorphism covariant.  In the simple example almost-associative geometry that we construct in section \ref{sec4}, the finite Dirac operator $D_{F}$ vanishes, so the Higgs fluctuations also vanish. We further discuss the split in equation  \eqref{GHsplit} in a later paper \cite{FarnsworthBoyle} where we explore models with non-trivial Higgs fields. 
 
As in regular gauge theory, we can determine the general form that fluctuations take from the form of the inhomogeneous gauge and higgs terms, i.e. $D\rightarrow D_A = D + B$ (except that now $B$ includes
both gauge and Higgs pieces). Under inner automorphisms of the input algebra, $\ast$-automorphism
covariance requires the fluctuations to transform as:
\begin{align}
 B&\rightarrow \hat{\alpha}B\hat{\alpha}^{-1} + \hat{\alpha}[D,\hat{\alpha}^{-1}].\label{Btrnsfm}
\end{align}
Thus the fluctuation $B$ ensures that the fluctuated Dirac operator $D_A$ transforms in the desired way under inner automorphisms, as given in equation \eqref{flucdirtran}.

For illustration let us give two examples.  In the first example, we determine the form that inner fluctuations take for the special case of an \textit{associative} almost-commutative geometry. As discussed in section \ref{Definitions} the inner automorphisms for an associative $\ast$-algebra $A$ are generated by elements of the algebra of derivations $\tilde{\delta}_c = L_{\tilde{c}} - R_{\tilde{c}} = c^i(L_{e_i} - R_{e_i})$, where the coefficients $c^{i}$ are real, and the algebra elements $e_{i}$ are anti-hermitian. In this case, equation \eqref{GHsplit} becomes
\begin{align}
  D'&\simeq  D - \underbrace{([D,c^k]{\tilde{e}_k}+ \epsilon'J[D,c^k]{\tilde{e}_k}J^*)}_{\text{Gauge field terms}}
 - \underbrace{(c^k[D,{\tilde{e}_k}]+ \epsilon'Jc^k[D,{\tilde{e}_k}]J^*)}_{\text{Higgs field terms}}\label{AssGHsplit}
\end{align}
where we have used the fact that $JD=\epsilon' DJ$ \cite{vandenDungen:2012ky}.  By inspection, the flucuated
Dirac operator should therefore be of the form:
\begin{align}
 D_A =D + B=D + A_{(1)} +\epsilon' JA_{(1)}J^* \label{ifassocfluc}
\end{align}
where $D$ is the un-fluctuated Dirac operator, and $A_{1}=\sum \tilde{a}[D,\tilde{b}]$ is a general one form.  The fluctuation $B$ is determined by the form of the derivation of an associative $\ast$-algebra $\tilde{\delta}_c = L_{\tilde{c}} - R_{\tilde{c}}$ (with $c$ anti-hermitian) along with the requirement that $B$ should be of order one ({\it i.e.}\ linear in $D$), and stable under fluctuations by the automorphisms of the (associative) algebra.
In this way, we recover the traditional formula for the 
fluctuation of the Dirac operator in the associative case.  
The `fluctuation' term $B$ is analogous to the connection term that appears in equation \eqref{covdiv}, and  is given by a one form with components valued in the algebra of derivations on $A$. Notice also that if $D$ acts as a derivation on the algebra representation and the Hilbert space, then the generalized one form
$A_{(1)}$ can be written as
\begin{align}
 A_{(1)} &= \sum L_{ \tilde{a}(D\tilde{b})}.\label{ifderassoc}
\end{align}
 
In the second example, we fluctuate the Dirac operator for an almost-associative geometry. Again the fluctuated Dirac opererator should transform under inner $*$-automorphisms of the input algebra as shown in equation \eqref{flucdirtran}. The only difference is that now the automorphisms will be generated by elements of the algebra of derivations $D(A)$ for the non-associative algebra in question, rather than by associative derivations of the form $\delta_{c}=L_{c}-R_{c}$.  Following the rest of the paper, the example we give is based on a finite spectral triple in which we represent the octonions on themselves. The octonions are an alternative algebra, and so their $*$-automorphisms will be generated by derivation elements for an alternative algebra. A general derivation will be given by an arbitrary sum of elements $\tilde{\delta}_{b,c} =
 [L_{\tilde{b}},L_{\tilde{c}}]+[L_{\tilde{b}},R_{\tilde{c}}]+[R_{\tilde{b}},R_{\tilde{c}}]\in D(A)$ (see equation \eqref{fmnonassocder}). To first order the Dirac operator must transform as
\begin{align}
  D' &\simeq D - [D,\tilde{\delta}_{b,c}]\nonumber\\
  &= D - [[D,L_{\tilde{b}}],L_{\tilde{c}}] +[[D,L_{\tilde{b}}],JL_{\tilde{c}}J^*] - \epsilon'J[[D,L_{\tilde{b}}],L_{\tilde{c}}]J^*
   \nonumber \\
    &\hspace{0.9cm}+ [[D,L_{\tilde{c}}],L_{\tilde{b}}] - \epsilon'[J[D,L_{\tilde{c}}]J^*,L_{\tilde{b}}]
 + \epsilon'J[[D,L_{\tilde{c}}],L_{\tilde{b}}]J^*, \label{nonaccderies}
\end{align}
where comparison between equations \eqref{AssGHsplit} and \eqref{nonaccderies} should be stressed. Once again the form of the fluctuated Dirac operator is determined by inspection of the inhomogeneous fluctuation terms and is given in the form $D_a = D + B$, where $D$ is the un-fluctuated Dirac operator, and the fluctuation term $B$ is given by:
\begin{align}
  B=\sum \delta_{A_{(1)},A_{(0)}}  :=\sum[A_{(1)},A_{(0)}] -[A_{(1)},JA_{(0)}J^*] + \epsilon'J[A_{(1)},A_{(0)}]J^*,\label{alflucnew}
\end{align}
where the sum is taken over generalized hermitian `one forms' $A_{(1)}$, and generalized 
`zero forms' $A_{(0)}$.  In this case, the fluctuation $B$ is determined by the form of the derivation of an {\it alternative} $\ast$-algebra $\tilde{\delta}_{a,b} = [L_{\tilde{a}},L_{\tilde{b}}]+[L_{\tilde{a}},R_{\tilde{b}}]+[R_{\tilde{a}},R_{\tilde{b}}]$, along with the requirement that $B$ should be of order one ({\it i.e.}\ linear in $D$), and stable under fluctuations by the automorphisms of the (alternative) $\ast$-algebra.  The `zero forms' $A_{(0)}$ will simply be given by left acting elements of the alternative algebra. The generalized `one forms' will depend on the representation of the algebra $\pi$, the real structure $J$, and the form of the un-fluctuated Dirac operator $D$. In the important case where $D$ acts as a derivation on the input algebra and Hilbert space
however, we have in comparison with equation \eqref{ifderassoc}
\begin{align}
 A_{(1)} = \sum L_{\tilde{a}(D\tilde{b})},\hspace{0.5cm}a,b\in\mathcal{A}.\label{oneformgen}
\end{align}
Again, these fluctuation terms transform as in equation \eqref{Btrnsfm}.  All associative algebras are also alternative algebras, and so the fluctuations given in equation \eqref{alflucnew} are a generalization of the fluctuations given in equation \eqref{ifassocfluc} for the associative case.  In particular, in the standard associative case ({\it i.e.}\ when the input algebra $A$ is associative, and the standard associative "order one" condition $[A_{(1)},J A_{(0)} J^{\ast}]=0$ holds), the central term on the right-hand-side of \eqref{alflucnew} vanishes, and \eqref{alflucnew} reduces to the associative expression \eqref{ifassocfluc}.  
 
Finally, in summarizing this section, let us return to recap a few important, but potentially confusing 
points.  Although the fluctuation of $D_{A}$ involves algebra elements $a\in A$ drawn from the 
non-associative algebra $A$, $D_A$ is simply a linear operator on $H$, and is not in any sense non-associative.  In particular, note that the fluctuations of $D$ are built not from the elements 
$a\in A$ themselves, but from $L_{a}$ and $R_{a}$, {\it i.e.}\ the (associative) operators which 
represent the left-action and right-action of $a$ on $H$.  Furthermore, these operators 
$L_{a}$ and $R_{a}$ are grouped together in a particular way, structured by the derivations of 
$A$.  This is ultimately done in order to ensure the $\ast$-automorphism covariance of the 
whole formalism.  We also remind the reader that, even when $A$ is non-associative, its 
automorphisms still form an ordinary (associative) group, and its derivations (from which 
the fluctuations of $D_{A}$ are built) still form an ordinary Lie algebra.  This means 
that, when we take an almost-associative geometry, and plug $D_{A}$ into the spectral
action, the spectral action will yield an ordinary Yang-Mills theory, just as it does in the 
almost-commutative case.  Let us now look at a concrete example.
 
\section{The Simplest Almost-Associative Example}
\label{sec4}

To illustrate and clarify the ideas introduced in Section \ref{NAGformalism}, in this section we will present the 
simplest example almost-associative geometry, based on the octonions and giving rise to a $G_2$ 
gauge theory via the spectral action.  We start by outlining the finite spectral data, and then explain 
how to fluctuate $D$ and compute the corresponding spectral action.

\subsection{The spectral triple}
\label{twisttrip}

We will consider the finite non-associative spectral triple given by
\begin{equation}
  F=\{\mathcal{A}_{F},\mathcal{H}_{F},D_{F},\gamma_{F},J_{F}\}
  =\{\mathbb{O},\mathbb{O},0,\mathbb{I},J_{\mathbb{O}}\}
\end{equation}
where $\mathcal{A}_{F}=\mathbb{O}$ is the octonion algebra, $\mathcal{H}_{F}=\mathbb{O}$
is the Hilbert space of octonions (equipped with its standard inner product), $\gamma_{F}=\mathbb{I}$ is the 
identity operator on $\mathcal{H}_{F}=\mathbb{O}$, and $J_{\mathbb{O}}$ denotes the ordinary octonionic conjugation operation on $\mathbb{O}$ \cite{2001math......5155B}.  The left and right action of 
$\mathcal{A}_{F}=\mathbb{O}$ on $\mathcal{H}_{F}=\mathbb{O}$ is the obvious one inherited from $\mathbb{O}$: we can always simply reinterpret an algebra -- even a non-associative one -- as bimodule over 
itself.  Note that, once we choose $\gamma_{F}=\mathbb{I}$, the choice $D_{F}=0$ is forced upon us by the
requirement that $\{D_{F},\gamma_{F}\}=0$ (see subsection \ref{spectral_triple}).  Then we can easily check that $J_{F}$ squares to unity, and commutes with $D_{F}$ and $\gamma_{F}$, which implies that the $KO$ dimension of the spectral triple is zero (see {\it e.g.}\ Sec. 2.2 in \cite{vandenDungen:2012ky}).
Because $D_{F}=0$, when we calculate the corresponding spectral action below, we will obtain a model in which there are no Higgs fields and the fermion fields are massless.  Again, this is just a consequence of the fact that we are considering the simplest non-associative spectral triple: in a forthcoming paper we will show how to construct more realistic models with Higgs fields, spontaneous symmetry breaking and fermion masses \cite{FarnsworthBoyle}.

Next we can construct the ``almost-associative geometry'' $T_c\times T_F$: {\it i.e.}\ the product of the canonical Riemannian spectral triple $T_c$ with the finite-dimensional nonassociative triple $T_F$ (see Subsection \ref{spectral_triple}):
\begin{equation}
  T_c\times T_F=\{C_{\infty}({\cal M},\mathbb{O}),
  L^{2}({\cal M},S)\otimes\mathbb{O},D_{c}\otimes\mathbb{I},
  \gamma_{c}\otimes\mathbb{I},J_{c}\otimes J_{\mathbb{O}}\},\label{allassg}
\end{equation}
where $D_c = -i\gamma^\mu\partial_\mu$. Note that $\mathbb{O}$ is an algebra over $\mathbb{R}$, so we take the
tensor product $C_{\infty} ({\cal M},\mathbb{R})\otimes \mathbb{O}$
over $\mathbb{R}$, as in \cite{2011IJGMM..08.1833D,
  2012arXiv1209.4832C}, to obtain $C_{\infty}({\cal M},\mathbb{O})$,
the algebra of smooth functions from ${\cal M}$ to $\mathbb{O}$.

\subsection{Fluctuating {$D$}}
\label{calcfluc}
The first task in constructing the spectral action for our almost-associative geometry is to fluctuate the Dirac operator $D=D_c\otimes\mathbb{I}$.  In our example, this task is simplified by the fact that $D_{F}=0$: this means the fluctuation will produce gauge fields, but no Higgs fields.  

As explained in Section \ref{NAGformalism}, we fluctuate $D$ by asking what must 
be added to it in order to make the spectral triple transform covariantly with respect
to automorphisms of $A$.  The automorphism group of the $\ast$-algebra $A=C_{\infty}
({\cal M},\mathbb{O})$ is the semi-direct product of two pieces: the group of outer 
automorphisms ${\rm Diff}({\cal M})$, and the group of inner automorphisms 
(the "gauge group" of maps from ${\cal M}$ to the automorphism group of $\mathbb{O}$).  
As compared to the story for an ordinary (associative) almost-commutative geometry 
 \cite{vandenDungen:2012ky}, there is nothing new about the outer automorphisms 
(which require the formalism to be covariant with respect to diffeomorphism, and give
rise to Einstein gravity via the spectral action in the usual way) so we focus here on the 
inner automorphisms.  Since $\mathbb{O}$ is an alternative algebra, its automorphisms
are generated by derivations of the form (\ref{fmnonassocder}); in particular, if we choose
any two of the seven "imaginary" octonionic basis vectors, ${\bf e}_{i}$ and ${\bf e}_{j}$,
we obtain a non-vanishing derivation: $\delta_{i,j}=[L_{{\bf e}_{i}},L_{{\bf e}_{j}}]+
[L_{{\bf e}_{i}},R_{{\bf e}_{j}}]+[R_{{\bf e}_{i}},R_{{\bf e}_{j}}]$.  Although there are 
$(7\times 6)/2 = 21$ such derivations, only 14 are linearly independent, and together 
they form the 14-dimensional exceptional Lie algebra $g_{2}$ \cite{RSchaf, 2001math......5155B}.
Let us call the 14 independent generators $\delta_{k}$ ($k=1,\ldots,14$): each is simply 
an $8\times8$ matrix from $\mathbb{O}\to\mathbb{O}$, and hence from $\mathcal{H}_{F}\to
\mathcal{H}_{F}$.  Thus, the inner automorphisms of $A$ are generated by derivations of 
the form ${\delta}=c^{k}(x)\otimes{\delta}_{k}$, where the coefficients $c^{k}(x)$ are arbitrary real 
functions on $\mathcal{M}$.  Let us now fluctuate the Dirac operator to account for these inner
automorphisms. Starting with the unfluctuated Dirac operator $D =
-i\gamma^\mu\partial_\mu\otimes\mathbb{I}$ we have, following section
\ref{flucDir}:
\begin{align}
  D\to D'={\rm e}^{\tilde{\delta}}D
                {\rm e}^{\tilde{\delta}}\approx D-[D,\tilde{\delta}]
                =-i\gamma^{\mu}(\partial_{\mu}\otimes\mathbb{I}
                -[\partial_{\mu},c^{k}(x)]\otimes\tilde{\delta}_{k}).
\end{align}
The inhomogeneous terms tell us the form that our general fluctuations should
take. The Dirac operator with inner fluctuation terms is given by:
\begin{align}
  D_{A}=-i\gamma^{\mu}[\partial_{\mu}\otimes\mathbb{I}
  +B_{\mu}^{k}(x)\otimes\tilde{\delta}_{k}], \label{frstord}
\end{align}
where $B_{\mu}^{k}(x)\otimes\delta_{k}$ is nothing but an ordinary 
Yang-Mills gauge field -- and, in particular, a $G_{2}$ gauge field -- written 
in slightly unfamiliar notation.  As mentioned above, we have focused here
on inner fluctuations, but the full Dirac operator must also be covariant
with respect to the outer automorphisms of the algebra as well -- {\it i.e.}\ we must
restore the spin connection.  Thus we have
\begin{align}
D_A =
-i\gamma^\mu\triangledown_\mu^E \label{fullfluc}
\end{align} where $\triangledown_\mu^E =
\triangledown_\mu^S \otimes \mathbb{I} + B_{\mu}^{k}(x)\otimes \tilde{\delta}_{k}$,
and $\triangledown_\mu^S$ is the usual Levi-Civita connection on the 
spinor bundle \cite{vandenDungen:2012ky}.

\subsection{The spectral action}
\label{calcboson}

Now that we have constructed the fluctuated Dirac operator $D_{A}$, we need no longer concern ourselves with the non-associativity of $A$ -- it has already played its role in shaping the bi-representation of $A$ on $H$, and hence the number of fermion fields, and the type and form of the bosonic fields we must add to make $D_{A}$ covariant.
From this point on, as far as evaluating the spectral action is concerned, all we need to 
know is that $H$ is an ordinary Hilbert space, and $D_{A}$ is an ordinary linear operator
on $H$ (and, in particular, a linear operator of the form such that the formalism of 
Section 3 in \cite{vandenDungen:2012ky} and, in particular, their Theorems 3.3 and
and 3.7 directly apply).

To be very explicit, we present here the resulting spectral action.  Calculating the spectral action corresponding to an operator $D_{A}$ of the form given in equation \eqref{fullfluc} is covered extensively in the literature, and we recommend {\it e.g.}\ the review given in Section 3 of \cite{vandenDungen:2012ky}, whose notation we follow here. Again, we stress there is nothing non-associative about the Dirac operator \eqref{fullfluc}. The spectral action is
given in terms of this $D_A$ as:
\begin{align}
\label{HKexp}
S_b = Tr\left(f(\frac{D_A}{\Lambda})\right)
\end{align}
where $f$ is a real, even function. Before we can perform the heat
kernel expansion we first need to calculate the square of the
fluctuated dirac operator, which is given by
\begin{align}
 D_A^2 &= (-i\gamma^\mu\triangledown_\mu^S\otimes \mathbb{I} -i\gamma^\mu
 B_{\mu}^{k}\otimes\tilde{\delta}_{k})^2\nonumber\\
&= \bigtriangleup_A^E -
\tfrac{1}{2}\gamma^\mu\gamma^\nu\otimes F_{\mu\nu} -
\tfrac{1}{4}R\otimes\mathbb{I},
\end{align}
where $R$ is the Ricci scalar, $B_{\mu}=B_{\mu}^{k}\delta_{k}$, and 
\begin{align}
 \bigtriangleup^E_A &=-g^{\mu\nu}\nabla_{\mu}^{E}\nabla_{\nu}^{E} \\
 F_{\mu\nu}&=\partial_{\mu}B_{\nu}-\partial_{\nu}B_{\mu}
 +[B_\mu,B_\nu].
\end{align}
Equation
\eqref{HKexp} can then be expanded as
\begin{align}
Tr\left(f(\tfrac{D_A}{\Lambda})\right)
&= 2f_4\Lambda^4a_0(D_A^2) + 2f_2\Lambda^2a_2(D_A^2) + f(0)a_4(D_A^2) +
O(\Lambda^{-2})
\end{align}
where the $f_{n}=\int_{0}^{\infty}f(x)x^{n-1}dx$ ($n>0$) and
$a_{k}(D_{\ast}^{2})$ are the Seeley-deWitt coefficients.  For a
compact Euclidean manifold without boundary we have
\begin{align}
a_0(D_A^2) &= \int_M d^{4}x\sqrt{g}\frac{1}{4\pi^2}\\
a_2(D_A^2) &= \int_M d^{4}x\sqrt{g}\frac{1}{48\pi^2}R\\
a_4(D_A^2) &= \int_M d^{4}x\sqrt{g}\frac{1}{16\pi^2}\frac{1}{360}{\rm Tr}[(\tfrac{5}{4}R^2 - 2R_{\mu\nu}R^{\mu\nu} + 2R_{\mu\nu\rho\sigma}R^{\mu\nu\rho\sigma}\nonumber\\
&\hspace{2.5cm} + 45\gamma^\mu\gamma^\nu\gamma^\rho\gamma^\sigma
F_{\mu\nu}F_{\rho\sigma}+
30\Omega_{\mu\nu}^E(\Omega^E)^{\mu\nu}],
\end{align}
where $\Omega_{\mu\nu}^E = \Omega_{\mu\nu}^S\otimes\mathbb{I} +
\mathbb{I}_4\otimes F_{\mu\nu}$, and $
{\rm Tr}(\Omega_{\mu\nu}^S\Omega^{S\mu\nu})
=-\tfrac{1}{2}R_{\mu\nu\rho\sigma}R^{\mu\nu\rho\sigma}$. The full
bosonic action is then
\begin{align}
S_b &\simeq
\int_M d^{4}x\sqrt{g}\frac{8}{(4\pi)^2}[
8f_4\Lambda^4 + \frac{2}{3}Rf_2\Lambda^2  \nonumber\\
&+\frac{f(0)}{360}(5R^2 - 8R_{\mu\nu}R^{\mu\nu} -7R_{\mu\nu\rho\sigma}R^{\mu\nu\rho\sigma} -\frac{240}{8}{\rm Tr}(F_{\mu\nu}F^{\mu\nu}))]
\end{align}
where we have used the fact that the finite Hilbert space has dimension $N=8$. To the bosonic spectral action we add the fermionic terms given by:\label{calcfermion}
\begin{align}
S_f =\langle\psi|D_A |\psi\rangle = \int \psi_i^\dagger(x)D_{A}^{ij} \psi_j(x)\sqrt{g}d^4x,
\end{align}
where $D_{A}^{ij} = -i\gamma^\mu(\triangledown_\mu^S\delta^{ij}
+ B_\mu^k\tilde{\delta}_k^{ij})$, is hermitian.  Note that, in this
equation, we are displaying explicitly the indices corresponding to the finite (8-dimensional)
Hilbert space ($i,j=1,\ldots,8$); the first $\delta^{ij}$ denotes an ordinary Kronecker delta,
while the second $\tilde{\delta}^{ij}_{k}$ are the previously discussed $G_{2}$ generators,
now with their $8\times8$ indices displayed.  In our convention the generators
$\tilde{\delta}_k^{ij}$ are anti-hermitian, which means that the gauge fields 
$B_\mu^k$ are hermitian.

The full action of our theory is given by the sum of both the bosonic
action and the fermionic action $S = S_b + S_f$. It describes Einstein
gravity coupled to a $G_2$ gauge theory, with 8 massless Dirac
fermions which split into a singlet and a septuplet under $G_2$. In
this paper, we have just presented the simplest possible model by way
of illustration.  In a forthcoming paper \cite{FarnsworthBoyle}, we
show how to construct more realistic physical models that include Higgs fields,
spontaneous symmetry breaking and fermion masses.

\appendix

\section{Twisted Geometry}
\label{Twistgeom}
In the previous section we constructed what is in some sense the
simplest finite non-associative geometry $T_F$ and used it to form an almost-associative geometry 
corresponding to a G2 gauge theory coupled to gravity. Its
simplicity however was not the only reason that we chose the octonion
example. It turns out that one may obtain our example finite
non-associative geometry through a `twisting' of an appropriate 
{\it associative} finite spectral triple. One can therefore arrive at
our example nonassociative spectral triple $T_F$ and check that it makes
sense, in two different ways.  On the one hand, $T_F$ satisfies all of
the required axioms for a
spectral triple (including the appropriate nonassociative
generalizations of the order zero and order one conditions presented
in Subsection \ref{realstrucsec}), and is compatible with the
principle of automorphism covariance, as explained in Subsection
\ref{autocov}.  On the other hand, one can start with an appropriate associative
spectral triple that satisfies the standard axioms for an associative spectral triple of $K0$ dimension zero, and then perform a so called `twist' into our nonassociative triple $T_F$. In this subsection we explain this twisting procedure.

We begin by introducing a few pieces of mathematical background. The octonions have a so called `quasialgebra' structure. 
For our present purposes a quasialgebra can be thought of as an
algebra that is, in some well defined way, related to certain other
algebras. Specifically, starting with an associative
quasialgebra $(\mathcal{A},\cdot)$, we can perform what is known as a
`twist' to obtain a new quasialgebra $(\mathcal{A}_F,\times)$. The new
algebra $\mathcal{A}_F$ shares the same underlying vector space as
$\mathcal{A}$ but has a new product (``$\times$'' instead of ``$\;\cdot\;$''). It is possible in this
way to describe the non-associativity of a quasialgebra
$(\mathcal{A}_F,\times)$ as resulting from a `twist' from an
associative quasialgebra $(\mathcal{A},\cdot)$.  

The authors Albuquerque and Majid
\cite{Albuquerque1999188} have already described in full detail the
octonions as a quasialgebra resulting from a `twist' on a particular associative group algebra. A group algebra is defined by taking the a group $G$ and its field $\mathbb{K}$
together in a natural way: namely, arbitrary linear
combinations of the form $\sum_{i}k_{i}g_{i}$, where
$k_{i}\in\mathbb{K}$ and $g_{i}\in G$.  These elements may be added
and multiplied in the obvious way, and thus form an algebra over the
field $\mathbb{K}$; the dimension of the algebra $\mathbb{K}G$ is just
the order of the group $G$.  $\mathbb{K}G$ is naturally a
$\ast$-algebra, with the $\ast$ operation given by
$(\sum_{i}k_{i}g_{i})^{\ast}= \sum_{i}k_{i}^{\ast}g_{i}^{-1}$; and it
is also naturally a Hilbert space, with the inner product of two
vectors $v^{(1)}=\sum_{i}k_{i}^{(1)}g_{i}$ and $v^{(2)}=\sum_{i}
k_{i}^{(2)}g_{i}$ given by $\langle v^{(1)}|v^{(2)}\rangle
=\sum_{i}(k_{i}^{(1)\ast})k_{i}^{(2)}$. 

In the case of the octonions, the corresponding associative group algebra of interest is $\mathbb{K}G$, where $\mathbb{K} = \mathbb{R}$, and  $G =
\mathbb{Z}_2\times\mathbb{Z}_2\times\mathbb{Z}_2$, so that $\mathbb{K}G$ is an 8-dimensional
algebra over the real numbers \cite{Albuquerque1999188, 2004JMP....45.3883A, MajFound, MajPrim}.  We can write each basis element of $\mathbb{K}G$ in the form $g_i =
(i_1, i_2, i_3)$, where $i_j\in \{0,1\}$; and then $\mathbb{K}G$
simply inherits the group multiplication law: $j \cdot k$ simply means
adding the two vectors (j and k), mod 2.  From here, we can obtain the
octonions by performing a `twist' -- i.e. by replacing the
multiplication law $x \cdot y$ with the new multiplication law:
\begin{align}
g_i\times g_j = g_i\cdot g_j F(g_i,g_j), \hspace{1cm} \forall g_i,g_j\in G\label{twist}
\end{align}
where $F$ is known as a `2-cochain twist' taking values in the field
$\mathbb{K}$ over which the algebra $\mathcal{A}_F$ is defined. The
2-cochain $F$ is given in our case as \cite{Albuquerque1999188}
\begin{align}
F(g_i,g_j) &= (-1)^f,\nonumber\\
f&=i_1(j_1+j_2+j_3) + i_2(j_2+j_3) + i_3j_3 + j_1i_2i_3 +i_1j_2i_3 + i_1i_2j_3. \label{Fvalue}
\end{align}
In discussing the twist from $\mathcal{A}=\mathbb{K}G$ to
 $\mathcal{A}_F=\mathbb{O}$ the authors Albuquerque and Majid
 \cite{Albuquerque1999188} give a `natural involution' ($*$ operation)
 on the twisted algebra basis
\begin{align}
Je_i = F(e_i,e_i)e_i
\end{align}
From equation \eqref{Fvalue} it can be seen that this involution is
simply octonionic conjugation. Prior to twisting we can simply take
$F(e_j,e_i) = 1$, $\forall e_i\in \mathbb{K}G$. Notice that in
$\mathbb{K}G$ each basis element is its own inverse. For this reason
the `natural' $*$ operation coincides in the untwisted case with what is
known as the `antipode' operator $S$ on $\mathbb{K}G$:
\begin{align}
Je_i = Se_i = e_i^{-1}
\end{align}
We can consider the data $\{A,H,J\} =
\{\mathbb{O},\mathbb{O},J_F\}$ as being `twisted' from the data
$\{\mathbb{K}G,\mathbb{K}G,S\}$. It is therefore natural to
consider a spectral triple $\{A, H,D,\gamma,J\}$ where
$A$ and $H$ are both given by $\mathbb{K}G$, and $A$ is represented in the obvious way: {\it i.e.}\ $\pi$ is the
identity map (so $\tilde{a}=a$), and the action of the operator
$\tilde{a}$ on an element of $H$ is given by the ordinary
product in $\mathbb{K}G$.  Furthermore, we can take $\gamma=1$; the
condition $\{\gamma,D\}=0$ then implies $D=0$.  Finally, the action of
$J$ on $H$ is naturally given by the $\ast$-operation in
$\mathbb{K}G$: $Jh=J(\sum_{i}k_{i}g_{i})=(\sum_{i}k_{i}g_{i})^{\ast}
=\sum_{i}k_{i}^{\ast}g_{i}^{-1}$. The twist given in equation \eqref{twist} then maps between the associative finite spectral triple corresponding to the group algebra $\mathbb{K}G$ and the non-associative finite spectral triple corresponding to the octonion algebra $A_F = \mathbb{O}$.

We are now in a position to analyze how the order zero condition
behaves under a `twist' from the associative $A =
H = \mathbb{K}G$ to the non-associative $A_F =
H_F = \mathbb{O}$. As $A$ is associative it will
satisfy the order zero condition given in
\eqref{order_zero_condition}.
\begin{align}
[\pi_{g_j}^0,\pi_{g_i}]\tilde{g_k} &=
(\tilde{g}_i\cdot \tilde{g}_k)\cdot\tilde{g}_j - \tilde{g}_i\cdot(\tilde{g}_k\cdot\tilde{g}_j) = 
0,\hspace{1cm}g_i,g_j,g_k\in G,
 \nonumber\\
\text{`twist'}\rightarrow
 0&=F^{-1}(g_i,g_k)F^{-1}(g_i\cdot g_k,g_j)(\tilde{g}_i\times \tilde{g}_k)\times \tilde{g}_j \nonumber\\
 &- F^{-1}(g_i,g_k\cdot g_j)F^{-1}(g_k,g_j)\tilde{g}_i\times (\tilde{g}_k\times \tilde{g}_j)\nonumber\\
&=\tfrac{F(g_i,g_k\cdot g_j)F(g_k,g_j)}{F(g_i,g_k)F(g_i\cdot g_k,g_j)}(\tilde{g}_i\times \tilde{g}_k)\times \tilde{g}_j -
\tilde{g}_i\times (\tilde{g}_k\times \tilde{g}_j)\nonumber\\
&=
\Phi_{\tilde{g}_i,\tilde{g}_k,\tilde{g}_j}^{-1}(\tilde{g}_i\times \tilde{g}_k)\times \tilde{g}_j
-
\tilde{g}_i\times (\tilde{g}_k\times \tilde{g}_j)\nonumber\\
&:= [R_{\tilde{g}_j},L_{\tilde{g}_i}]_\Phi \tilde{g}_k\label{oldzero}
\end{align}
where the `associator' is defined as
$\Phi_{\tilde{g}_i,\tilde{g}_k,\tilde{g}_j}:=
\tfrac{F(g_i,g_k)F(g_i\cdot g_k,g_j)}{F(g_i,g_k\cdot
  g_j)F(g_k,g_j)}$. After the twist we should consider the basis
elements $g_i,g_j\in A_F$ and $g_k\in H_F$. Equation
\eqref{oldzero} suggests we introduce an augmented order zero
condition in the general sense given by
\begin{align}
[R_{\tilde{b}},L_{\tilde{a}}]_\Phi =0\hspace{1cm}\forall a,b \in A_F. \label{orderz}
\end{align}
Here the subscript $\Phi$ can be seen as telling us to `flip' the
brackets on one side of the commutator when acting on a hilbert space
element. Note that for an associative algebra, the `associator' $\Phi$
will be trivial and our augmented order zero condition will collapse back to
that given in the associative case \eqref{order_zero_condition}. Note also, that for our octonion example, when $a = b$, the  `associator' $\Phi$ will again be trivial, as would be expected following the conditions given in equation \eqref{altor01}

We would like to stress that we can arrive at the nonassociative
spectral triple $T_F$, and check that it makes sense, in two different
ways.  On the one hand, $T_F$ satisfies all of the required axioms for a
spectral triple (including the appropriate nonassociative
generalization of the order zero condition presented
in Subsection \ref{realstrucsec}), and is compatible with the
principle of automorphism covariance, as explained in Subsection
\ref{autocov}.  On the other hand, we can start with the associative
spectral triple: $T_{F_{0}}=\{\mathbb{K}G,\mathbb{K}G,
0,\mathbb{I},J_{\mathbb{K}G}\}$, where $\mathbb{K}G$ is the group
algebra based on $\mathbb{K}=\mathbb{R}$ and
$G=\mathbb{Z}_{2}\times\mathbb{Z}_{2} \times\mathbb{Z}_{2}$, and
$J_{\mathbb{K}G}$ denotes the natural $\ast$ operation in
$\mathbb{K}G$ (see Subsection \ref{spectral_triple}). This spectral
triple satisfies the
standard axioms for an associative spectral triple of $K0$ dimension
zero.  But then, when one twists $\mathbb{K}G$ into $\mathbb{O}$ (see
\cite{Albuquerque1999188}), the associative spectral triple $T_{F_{0}}$ is
correspondingly twisted into our nonassociative triple $T_F$.

\acknowledgments

It is a pleasure to thank Florian Girelli for his advice on early drafts of this paper. SF would also like to thank Nima Doroud for his time in useful discussion. This work is supported by the Perimeter Institute for Theoretical Physics. Research at the Perimeter Institute is supported by the Government of Canada through Industry Canada and by the Province of Ontario through the Ministry of Research \& Innovation.  LB also acknowledges support from a Discovery Grant from the Natural Sciences and Engineering Research Council of Canada.

\bibliographystyle{JHEP}   
\bibliography{mybib} 

\end{document}